\documentclass[fleqn,usenatbib]{mnras}
\usepackage{newtxtext,newtxmath}
\usepackage{units}
\usepackage[T1]{fontenc}
\usepackage{anyfontsize} % suppression of font size warning
\usepackage{xcolor}
\usepackage{calc}

\DeclareRobustCommand{\VAN}[3]{#2}
\let\VANthebibliography\thebibliography
\def\thebibliography{\DeclareRobustCommand{\VAN}[3]{##3}\VANthebibliography}

\usepackage{graphicx}	% Including Fig. files
\graphicspath{{figures}}

\title[Changing-Look Inspirals]{Changing-Look Inspirals: Trends and Switches in AGN Disk Emission as Signposts for Merging Black Hole Binaries}

\author[Zrake et al.]{
Jonathan Zrake$^{1}$\thanks{E-mail: jzrake@clemson.edu}
Madeline Clyburn,$^{1}$
Samuel Feyan$^{1}$ \\
$^{1}$Department of Physics and Astronomy, Clemson University, Clemson, SC 29634, USA\\
}

\date{Accepted XXX. Received YYY; in original form ZZZ}

\pubyear{2024}

\begin{document}

\label{firstpage}
\pagerange{\pageref{firstpage}--\pageref{lastpage}}
\maketitle

\begin{abstract}
Using grid-based hydrodynamics simulations and analytic modeling, we compute the electromagnetic (EM) signatures of gravitational wave (GW) driven inspirals of massive black hole binaries that accrete gas from circumbinary disks, exploring the effects of varying gas temperatures, viscosity laws, and binary mass ratios. Our main finding is that active galactic nuclei (AGN's) that host inspiraling binaries can exhibit two sub-types of long-term secular variability patterns: Type-A events which dim before merger and brighten afterward, and Type-B events which brighten before merger and dim afterward. In both types the merger coincides with a long-lasting chromatic change of the AGN appearance. The sub-types correspond to the direction of angular momentum transfer between the binary and the disk, and could thus have correlated GW signatures if the gas-induced torque can be inferred from GW phase drift measurements by LISA. The long-term brightness trends are caused by steady weakening of the disk-binary torque that accompanies orbital decay, it induces a hysteresis effect whereby the disk ``remembers'' the history of the binary's contraction. We illustrate the effect using a reduced model problem of an axisymmetric thin disk subjected at its inner edge to the weakening torque of an inspiraling binary. The model problem yields a new class of self-similar disk solutions, which capture salient features of the multi-dimensional hydrodynamics simulations. We use these solutions to derive variable AGN disk emission signatures within years to decades of massive black hole binary mergers in AGN's. Spectral changes of Mrk 1018 might have been triggered by an inspiral-merger event.
% We discuss prospects to extract the sign of the gas-induced torque from low-frequency GW observations with LISA.
% \the\columnwidth
% \the\textwidth
\end{abstract}

\begin{keywords}
accretion disks -- black hole physics -- black hole mergers -- hydrodynamics
\end{keywords}

\section{Introduction}
The gravitational wave (GW) driven inspirals of massive black hole binary systems ($M \sim 10^{4-7} M_\odot$) are promising targets for planned low-frequency GW detectors, including LISA, Taiji, and TianQin \citep{amaro-seoane_astrophysics_2023, ruan_taiji_2020, gong_concepts_2021}. Massive black hole binaries likely form in gas-rich environments \citep{sanders_luminous_1996, barnes_transformations_1996} and may accrete from circumbinary gas disks \citep{begelman_masive_1980}, creating binary active galactic nuclei (AGN's). If a binary in an AGN merges, it could produce an electromagnetic (EM) counterpart accompanying the GW-driven inspiral and merger \citep[e.g.][]{schnittman_electromagnetic_2011, bogdanovic_electromagnetic_2022, dorazio_observational_2023}. Such EM counterparts will be needed to match GW events with their host galaxies, because the error volume associated with GW source will generally contain many galaxies \citep{mangiagli_massive_2022}.

A promising category of binary merger EM signature may be found in long-term secular changes of AGN light. In contrast to periodic variability signatures, which are expected \cite[e.g.][]{dorazio_observational_2023,westernacher-schneider_multiband_2022}, but can suffer from relatively high false alarm rates due to stochastic AGN variability \citep[e.g.][]{liu_did_2018}, long-term secular changes may be more confidently identified in time-domain EM surveys \cite[e.g.][]{penil_multiwavelength_2024}. They may also be less prone to false negatives associated with finite observing cadences \citep[e.g.][]{xin_identifying_2024}.

AGN's can exhibit striking secular changes, such as changing-look events, where there is a lasting chromatic change of an AGN's appearance often in conjunction with the appearance or disappearance of broad-line spectral features \citep[e.g.][]{ricci_changing-look_2023}. Some changing-look events are known to be driven by a restructuring of the inner portions of the AGN disk \citep[e.g.][]{yang_probing_2023, veronese_interpreting_2024}, however no physical mechanism has yet emerged as being clearly favorable, and there could certainly be a variety of underlying causes.

Nevertheless there are good reasons to explore connections between binary mergers and secular changes of AGN appearances. In particular, AGN's have been proposed to become active in the time period following a merger, as gas from an extended disk around the binary flows onto the black hole merger remnant, following an extended quiescent phase \citep{liu_doubledouble_2003, milosavljevic_afterglow_2005, shapiro_filling_2010, tanaka_time-dependent_2010, tanaka_witnessing_2010}. Broadly, the post-merger light curves are predicted to grow brighter and bluer on timescales of anywhere from months \citep{tanaka_witnessing_2010} to millenia \citep{liu_doubledouble_2003}.

In those studies from between 2003 and 2010, the circumbinary disk is described as being truncated far outside the binary orbit \citep{pringle_accretion_1981} such that black hole accretion is delayed until after the merger. If instead the black holes were to be actively accreting from the circumbinary disk, then the AGN would be already active, and it might show correlated pre-merger and post-merger variability signatures. In binaries with significantly unequal masses, some ``fossil gas'' is envisioned to surround the larger black hole, and to be driven inwards by the inspiraling secondary, possibly leading to a luminous outflow in the late inspiral phase \citep{armitage_accretion_2002, lodato_black_2009, chang_fossil_2010}. 

However, binaries of all mass ratios are thought to accrete gas \cite[e.g.][]{artymowicz_mass_1996,macfadyen_eccentric_2008}, and multi-dimensional simulations indicate such accretion can be sustained until a late stage of a GW-wave driven inspiral \citep[e.g.][]{zanotti_electromagnetic_2012, noble_circumbinary_2012, farris_binary_2015, tang_late_2018, avara_accretion_2024}. Simulations also broadly indicate that black hole accretion is suppressed for a window of time around the merger \citep[e.g.][]{krauth_disappearing_2023, franchini_emission_2024}. This suppression results from the binary having ``viscously decoupled'' from the disk, meaning that the black holes have begun to spiral in faster than the disk viscously spreads inward to fuel the binary \citep{dittmann_decoupling_2023}. The viscous decoupling implies a characteristic timescale $t_{\rm dec}$, which corresponds to the window of time, containing the merger, during which the black hole accretion is significantly suppressed. We estimate that for LISA sources $t_{\rm dec}$ could be as short as hours to weeks, but it is sensitive to the uncertain magnitude of the disk internal stress.

Recently \cite{clyburn_dynamics_2024} reported simulations of inspiraling binaries where the accretion rates were noticeably reduced (at the level of $\sim 10\%$) even many hundreds of $t_{\rm dec}$ before the merger. It was suggested in that paper that such long-term temporal effects could implicate a long-range or hysteresis disk effect \cite[e.g.][]{rafikov_structure_2013, rafikov_accretion_2016}, in which the binary's past evolution is imprinted on the large-scale structure of the disk. Understanding such long-term effects could greatly enhance efforts to select host galaxy candidates based on AGN variability patterns.

In this paper we show that long-range disk effects are expected, and do cause predictable brightness trends in accreting GW-driven inspirals. The trends are caused by the gradual reduction in the magnitude of the disk-binary torque coupling; essentially, as the binary orbit shrinks, the disk is acted on by an object with a decreasing lever-arm. This ``weakening-torque effect'' leaves a lasting imprint on the radial distribution of mass in the accretion disk, and modulates the gas supply to the black holes over timescales much exceeding $t_{\rm dec}$ before and after the merger. The effect also implies that before the merger the AGN disk can grow brighter, or dimmer, depending on the direction of the angular momentum flow between the binary and the disk. The mass and angular momentum exchange also causes a drifting of the GW phase relative to that of a vacuum inspiral, and might be measurable in future GW observations \citep[e.g.][]{chakrabarti_gravitational_1996, levin_starbursts_2007, kocsis_observable_2011, derdzinski_probing_2019, derdzinski_evolution_2021}.

Specifically, we predict that binaries found via GW measurements to be torqued positively by gas will show concave-up bolometric light curves centered on the merger time (i.e. the AGN is dimmest around merger). We call these events Type-A for ``attenuating''. Whereas, binaries that are torqued negatively by gas will show concave-down bolometric light curves (brightest around merger), aside from within $t_{\rm dec}$ of the merger itself. We call these events Type-B for ``brightening'', and demonstrate they might also exhibit a fast-rising soft X-ray flare which decays over the years following the merger. We call these events ``changing-look inspirals'' or CLI's, because they all cause a lasting chromatic change in the AGN appearance coinciding with the GW burst.

Our results are based on new 2D hydrodynamics simulations of accreting binary inspirals, run for very long durations ($10^{4-5}$ orbits) before and after the merger. These simulations accurately gauge the accretion rate trends during the inspiral and post-merger phases. We include cases with different viscosity laws and binary mass ratios. We also report one 2D simulation where the binary is surrounded by a relatively cold disk, resulting in a negative binary torque \citep{tiede_gas-driven_2020, penzlin_binary_2022}, and confirm that it displays an upward trending pre-merger accretion rate, a significant spike shortly following the merger, and an extended downward trending post-merger accretion rate. We also show numerical integrations of the axisymmetric thin disk equations, where the disk is subjected at its inner edge to a time-varying torque accurate for a GW inspiral, and show that aside from stochastic variations in the 2D calculations, the 1D and 2D results are generally in good agreement. Finally, we show that the large-scale disk evolution tracks very closely a class of analytic functions of the form $(1 + x^{-k_1})^{k_2}$, where the positive constants $k_1$ and $k_2$ depend weakly on the radial distribution of the disk internal stresses, and $x$ can stand for either $t / t_{\rm dec}$ or $r / r_{\rm dec}$ depending on the context ($r_{\rm dec}$ is the binary separation at time $t_{\rm dec}$ before the merger). These functions are derived from a two-zone disk approximation, and encode the time series of the accretion luminosity as well as the disk radial structure. They also enable fast estimations of the observable EM and GW signatures of the weakening-torque effect.

Our paper is organized as follows. In Sec. \ref{sec:model} we describe the weakening-torque effect, and derive the time-dependent self-similar disk solutions. In Sec. \ref{sec:comparisons} we present new 2D hydrodynamics simulations of inspiraling binaries, and compare the accretion rate trends from the 2D simulations to 1D numerical simulations and to the analytic solutions. In Sec. \ref{sec:observables} we derive EM observables from the analytic solutions, including multi-band light curves and time-varying emission spectra. In Sec. \ref{sec:discussion} we discuss possible connections between binary mergers and changing-look AGN's including Mrk 1018. We also remark on prospects to infer the sign of the gas-induced binary torque from GW measurements. A summary is provided in Sec. \ref{sec:summary}. Throughout the paper, time is measured relative to the binary merger, i.e. $t<0$ during the inspiral.

\section{The Weakening-Torque Effect}\label{sec:model}
\subsection{Standard Disk Around a Binary}
We begin by describing a schematic model for the time evolution of a disk which is coupled at its inner edge to a binary undergoing GW driven orbital decay. The disk is geometrically thin, axisymmetric, and has a Keplerian rotation profile. The governing equation is \citep[e.g.][]{pringle_accretion_1981}
\begin{equation}
\frac{\partial}{\partial t} \Sigma(r, t) - \frac{3}{r}\frac{\partial}{\partial r}\left[ \frac{r}{l} \frac{\partial}{\partial r}  \left(\nu  \Sigma l\right) \right] = 0 \, ,
\label{eqn:standard_disk}
\end{equation}
where $\Sigma(r, t)$ is the vertically integrated mass density, $\nu(r)$ is the kinematic (turbulent) viscosity coefficient, and $l \equiv \sqrt{G M r}$ is the specific angular momentum of orbiting gas at radius $r$. Mass flows inwards through the large-scale disk at rate $\dot M_\infty$, and delivers mass to the black holes at rate $\dot M(t)$. We use the term ``disk'' to refer either to the pre-merger circumbinary disk, or to the post-merger disk surrounding the black hole merger remnant.

Before the merger, the disk is subjected at its inner edge to a time-varying torque, parameterized as $\dot J(t) \equiv \ell \dot M(t) \Omega a^2$, where $a$ is the (decreasing) binary semi-major axis, $\Omega \equiv \sqrt{G M / a^3}$ is the binary orbital frequency, and $\ell$ is called the torque parameter. It is set by complex gas flows in the near vicinity of the binary, and in general must be measured from multi-dimensional simulations. It can be positive or negative depending on the binary parameters and intrinsic gas properties \citep{tang_orbital_2017, munoz_hydrodynamics_2019, munoz_circumbinary_2020, duffell_circumbinary_2020, tiede_gas-driven_2020, zrake_equilibrium_2021, dorazio_orbital_2021, penzlin_binary_2022, siwek_orbital_2023}. Note that $\dot M(t)$ and $\dot J(t)$ are the mass and angular momentum currents flowing inwards through the inner edge of the disk, and that $\dot J(t)$ includes the gravitational tidal torque $\dot J_{\rm grav}$ exerted by the disk on the binary, as well as the accretion torque $\dot J_{\rm acc} \sim \dot M \Omega a^2$, which arises from the addition of co-orbiting gas to the black holes.

\subsection{The Two-Zone Disk Approximation}\label{subsec:two_zone}
An approximate time-dependent solution to Eqn. \ref{eqn:standard_disk} can be constructed from a two-zone heuristic picture \citep[e.g.][]{syer_satellites_1995}. We envision the disk to have a relaxed inner zone, and frozen outer zone. In the relaxed zone the disk adiabatically evolves through a sequence of steady-states, each reflecting the instantaneous binary torque. The surface density in the relaxed zone is thus
\begin{equation}
\Sigma_{\rm in}(r, t) = \frac{\dot M(t)}{3 \pi \nu(r)} \left(1 - \ell\sqrt{\frac{a(t)}{r}} \right) \, .
\label{eqn:sigma_in}
\end{equation}
The relaxed zone extends outwards to a critical radius $r_\nu(t)$, beyond which the disk relaxes more slowly than the orbit contracts; $r_\nu(t)$ is the radius for which the viscous relaxation time $t_{\rm visc}(r) \equiv 2 r^2 / 3\nu(r)$ equals the characteristic orbital contraction time $|a / \dot a|$. We call $r_\nu$ the radius of influence
\footnote{A radius of influence has been used in various contexts, see \cite{syer_satellites_1995, ivanov_evolution_1999, rafikov_structure_2013, rafikov_accretion_2016, duffell_santa_2024}. Those works treated $r_\nu$ as a length scale that increased over time, transmitting the binary's influence outward through the disk. We envision $r_\nu$ to have already moved far away from the binary in an earlier gas-driven evolutionary phase. In Sec. \ref{subsec:pre_merger_accretion_rate}, $r_\nu$ moves inwards as GW losses dominate the orbital decay. The radius of influence moves outwards again after the merger (Sec. \ref{subsec:post_merger_accretion_rate}).}.
At radius $r > r_\nu(t)$ the disk reflects the state of the binary at an earlier time $t_\nu(r)$ when the $r_\nu$ moved inwards across radius $r$. The surface density $\Sigma_{\rm out}(r)$ in the non-relaxed zone $r > r_\nu(t)$ is thus frozen at the time $t_\nu(r)$,
\begin{equation}
\Sigma_{\rm out}(r) \equiv \Sigma_{\rm in}(r, t_\nu(r)) \, .
\label{eqn:sigma_out_def}
\end{equation}
% We thus approximate the surface density there as frozen in time,
% Note that $t_\nu(r)$ differs from $t_{\rm visc}(r)$ by a constant.
% In the viscously non-relaxed zone, the disk changes little in the time remaining before the binary merges.

\begin{figure}
    \includegraphics[width=\columnwidth]{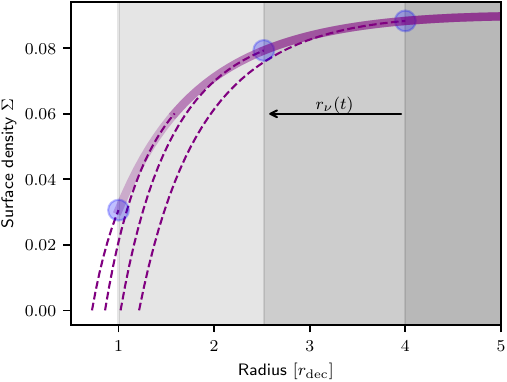}
    \caption{Illustration of the two-zone disk approximation. Thick solid lines show $\Sigma_{\rm out}$ at representative times $t/t_{\rm dec}=-8,-4,-2,-1$. Dashed lines show $\Sigma_{\rm in}$. Filled circles highlight the matching condition $\Sigma_{\rm in}(r_\nu(t), t) = \Sigma_{\rm in}(r_\nu(t), t - dt)$ which leads to Eqn. \ref{eqn:bc}. Vertical boundaries between grey shaded regions show the radius of influence $r_\nu(t)$. This model uses $n=1/2$ and $\ell=0.85$.}
    \label{fig:two_zone_model_diagram}
\end{figure}

A sequence of surface density profiles in the two-zone approximation is illustrated in Fig. \ref{fig:two_zone_model_diagram}. At time $t - dt$, the surface density within $r < r_\nu(t - dt)$ is given by $\Sigma_{\rm in}(r, t - dt)$ in Eqn. \ref{eqn:sigma_in}. At time $t$, the portion of the disk within $r_\nu(t)$ has been changed to $\Sigma_{\rm in}(r, t)$. Because the disk is frozen outside $r_\nu(t)$, the new surface density profile (at time $t$) matches the old surface density profile (at time $t - dt$) at the new radius of influence $r_\nu(t)$. The matching condition can thus be expressed as either $\Sigma_{\rm in}(r_\nu(t), t) = \Sigma_{\rm in}(r_\nu(t), t - dt)$, or as
\begin{equation}
\frac{\partial}{\partial t} \Sigma_{\rm in}(r, t)\big|_{r=r_\nu(t)} = 0 \, .
\label{eqn:bc}
\end{equation}
Given the equation for $\dot a(t)$ in a GW inspiral, and a choice of viscosity law $\nu(r)$, the evolution of $r_\nu(t)$ is then determined, and an ordinary differential equation for $\dot M(t)$ can be derived.

In addition to the time-evolving length scales $a$ and $r_\nu$, there is a third time-evolving length scale, $r_\star \equiv \ell^2 a$. At $r_\star$, gas parcels on circular orbits have specific angular momentum $|\dot J(t)| / \dot M(t)$. When $\ell > 0$, the surface density formally goes to zero at $r_\star$.

\subsection{Accretion Rate Trends Before Merger}\label{subsec:pre_merger_accretion_rate}
During the GW driven inspiral, the binary semi-major axis evolves according to the post-Newtonian inspiral \citep[e.g.][]{peters_gravitational_1964},
\begin{equation}
\frac{\dot a}{a} = -\frac{64 \eta c r_g^3}{5 a^4}
\quad
\left\{ r_g \equiv G M / c^2 \right\} \, ,
\label{eqn:adot}
\end{equation}
where $\eta \equiv M_1 M_2 / M^2$ is the symmetric mass ratio. As will be shown in a moment, $r_\nu$ decreases faster than $a$, and this means there is a characteristic timescale at which $r_\nu = a$. This timescale is called the viscous decoupling time and it is denoted as $t_{\rm dec}$. It corresponds to a length scale $r_{\rm dec}$, which is the binary separation at time $t_{\rm dec}$ before the merger. The solution of Eqn. \ref{eqn:adot} can be written as $a(t) = r_{\rm dec}(-t / t_{\rm dec})^{1/4}$, where $t$ increases through zero at the merger time. For a power-law viscosity profile $\nu(r) \propto r^n$, the radius of influence is
\begin{equation*}
r_\nu = r_{\rm dec} \left( \frac{a}{r_{\rm dec}} \right)^\beta
\quad
\left\{ \beta \equiv \frac{4}{2-n} \right\}
\, .
\end{equation*}
Use of the matching condition Eqn. \ref{eqn:bc} then leads to a first-order ordinary differential equation for $\dot M(t)$,
\begin{equation*}
\frac{1}{\dot M}\frac{d \dot M}{d x} = \frac{\ell / 2}{x^m - \ell x} 
\quad
\left\{x(t) \equiv a(t) / r_{\rm dec}
\quad
m \equiv \frac{n - 6}{2n - 4}
\right\}
\, ,
\end{equation*}
which has the exact solution
\begin{equation}
\dot M_{\rm pre}(t) \equiv \dot M_\infty \left[1 - \ell \left( \frac{-t}{t_{\rm dec}} \right)^{-p/8} \right]^{1/p}
\quad
\left\{ p \equiv \frac{2 + n}{2 - n} \right\}
\, .
\label{eqn:mdot_pre}
\end{equation}

The two-zone model thus predicts that the pre-merger accretion rate could trend upwards, or downwards, depending on the sign of the total binary torque. For the special case of $\alpha$-viscosity, $n = 1/2$, $\beta = 8/3$, $p = 5/3$, and Eqn. \ref{eqn:mdot_pre} has three free parameters: $\dot M_\infty$, a time constant $\tau = |\ell|^{24/5} t_{\rm dec}$, and the sign of the torque.
At times sufficiently larger than $\tau$, the quantity $\delta \equiv 1 - \dot M(t) / \dot M_\infty$ scales as $|t|^{-5/24}$. This result was mentioned in Sec. 4.2 of \cite{clyburn_dynamics_2024}. For an arbitrary viscosity law, the time constant is $\tau \equiv |\ell|^{8/p} t_{\rm dec}$. Note that Eqn. \ref{eqn:mdot_pre} is not expected to hold for $|t| \lesssim t_{\rm dec}$, because the disk around $r_\nu$ is then strongly influenced by the binary. However we estimate in Sec. \ref{subsec:decoupling_timescale} that for LISA sources, the viscous decoupling time is likely on the order of hours to weeks. We show in Sec. \ref{sec:comparisons} that for $\ell > 0$ the model agrees surprisingly well with 1D and 2D simulations even within $t_{\rm dec}$ of the merger, whereas for $\ell < 0$ the model fails some $t_{\rm dec}$ before and after the merger.

Having obtained the pre-merger accretion rate $\dot M_{\rm pre}(t)$, we can now derive the surface density profile in the frozen part of the disk $r > r_\nu(t)$. This is done by substituting Eqns. \ref{eqn:adot} and \ref{eqn:mdot_pre} into Eqn. \ref{eqn:sigma_in} and then invoking Eqn. \ref{eqn:sigma_out_def},
\begin{equation}
\Sigma_{\rm out}(r) = \frac{\dot M_\infty}{3 \pi \nu(r)} \left[ 1 - \ell \left( \frac{r}{r_{\rm dec}} \right)^{-q/2} \right]^{1/q} \quad \left\{ q \equiv \frac{2+n}{4} \right\} \, .
\label{eqn:sigma_out}
\end{equation}
This is the form of the surface density at the time of the merger, down to radius $r_{\rm dec}$. It means the binary merges in a relatively more dilute environment when $\ell > 0$, and in a denser environment when $\ell < 0$, corresponding to attenuating (Type-A) and brightening (Type-B) events respectively. Note that the surface density profile deviates significantly from a step-function, so the post-merger accretion trends will differ quantitatively from those in \cite{shapiro_filling_2010} and \cite{tanaka_time-dependent_2010}, where the viscous refilling was computed starting from a step-function surface density profile at $t=0$.

\subsection{Accretion Rate Trends After Merger}
\label{subsec:post_merger_accretion_rate}
Here we derive the accretion rate $\dot M_{\rm post}(t)$ onto the remnant black hole. This is done by constructing a new sequence of two-zone disks, analogous to the procedure in Sec. \ref{subsec:two_zone}, where now the radius of influence $r_\nu(t)$ moves outwards following the merger, as the disk relaxes around the merger remnant to a (nearly) zero-torque profile. The radius of influence is given implicitly by $t_{\rm visc}(r_\nu(t)) \equiv 4 \kappa t$ where $\kappa$ is an unknown constant of order unity, and the factor of $4$ is for convenience. The relaxed portion of the remnant disk has a small angular momentum current,
% \footnote{The remnant disk has a non-zero torque at its inner edge because the merger remnant has a finite size. It is small compared that of the binary, however the effect on the post-merger accretion rate could be straightforwardly computed. It would be sensitive to the mass and spin of the merger remnant. Also note that the approximation in Eqn. \ref{eqn:mdot_post} uses the same mass for the binary and the remnant black hole. Accounting for reduction of the central object gravitating mass by GW energy loss would slightly change the form of the post-merger accretion rate.}
so inside of $r_\nu(t)$, the surface density is given approximately by Eqn. \ref{eqn:sigma_in} with $\ell$ set to zero, and $\dot M(t) = \dot M_{\rm post}(t)$.

Outside the radius of influence, the disk is frozen, in the sense that the surface density has changed only very little since the time of the merger; outside of $r_\nu(t)$ then $\Sigma(r, t)$ is given by Eqn. \ref{eqn:sigma_out}. The post-merger accretion rate is found by matching $\Sigma_{\rm in}$ and $\Sigma_{\rm out}$ at the radius of influence, and solving for the accretion rate. We find
\begin{equation}
\dot M_{\rm post}(t) =  \dot M_\infty \left[1 - \ell \left(\frac{\kappa t}{t_{\rm dec}}\right)^{-p / 8} \right]^{1/q} \, .
\label{eqn:mdot_post}
\end{equation}
The constant $\kappa$ is determined empirically by fitting to the simulation results.

\subsection{How is Mass Conserved?}
It might be puzzling at first that in the case of $\ell < 0$ the accretion rate exceeds $\dot M_\infty$ both before and after the merger. Indeed one could reasonably expect that to conserve mass, the accretion rate would need to be suppressed after the merger if it had been enhanced before. The resolution to this ``paradox'' is simply that more mass has accumulated in the inner portion of the circumbinary disk when the torque is negative than when it is positive. This can be seen from the steady-state radial surface density profile in Eqn. \ref{eqn:sigma_in}. When the binary torque is negative, the enhanced accretion rate around the merger time comes from the withdrawal of mass that was saved up a long time ago, when the disk first relaxed around the binary (back then, the binary accreted more slowly while the pile-up was developing). As the torque weakens, the mass accumulated at the inner edge of the circumbinary disk is gradually released and flows onto the binary. The situation is reversed when the torque is positive.

\section{Comparison With Simulations}\label{sec:comparisons}
Here we present new 2D grid-based hydrodynamics simulations of accreting binaries in the inspiral and post-merger phases. We also present 1D simulations of a standard disk which is subjected to a time-varying torque at its inner edge. We compare the calculations from the 1D and 2D numerical settings to the two-zone model derived in Sec. \ref{sec:model}. We briefly describe the 1D and 2D codes in Sec. \ref{subsec:1d_code} and Sec. \ref{subsec:2d_code} respectively. Finally in Sec. \ref{subsec:results} we show detailed three-way comparisons for a set of three representative disk-binary configurations: two cases of positively torqued binaries with different viscosity laws, and one case of a negatively torqued binary.

\subsection{The 1D Code}\label{subsec:1d_code}
The 1D code solves Eqn. \ref{eqn:standard_disk} in a flux-conservative form using the method of lines,
\begin{equation*}
m_i^{n+1} = m_i^n + \left(f^n_{i+1/2} - f^n_{i-1/2} \right) \Delta t \, .
\end{equation*}
Here, $m_i^n$ is the mass within the annular zone with radial index $i$ at time level $n$, and $f_{i+1/2}$ is the (inward) mass flux through the $i+1/2$ zone interface. The mass flux is $f = 2 \pi r \varv_{\rm drift} \Sigma$, where the (inward) viscous radial drift speed is 
\begin{equation*}
\varv_{\rm drift} = \frac{3}{\Sigma l} \frac{\partial}{\partial r} \left( \nu \Sigma l \right) \, ,
% \label{eqn:drift_speed}
\end{equation*}
and as before $l$ is the specific angular momentum of orbiting gas at radius $r$. The drift speed at zone interface $i+1/2$ is obtained by a first-order difference of $\nu \Sigma l$ in zones $i$ and $i+1$.

The binary's effect on the disk is modeled as a time-dependent boundary condition applied at the domain inner edge. This procedure is similar in spirit to the one in \cite{pringle_properties_1991}, except we allow the torque to be either positive or negative following the ansatz $\dot J = \ell \dot M \Omega a^2$. This is accomplished starting with the general relation \citep[e.g.][]{rafikov_accretion_2016}
\begin{equation}
\frac{d l}{d r} f = \frac{\partial g}{\partial r}
\label{eqn:rafikov_viscous_torque}
\end{equation}
between the local mass flux $f$ and the viscous torque $g \equiv 3 \pi \nu \Sigma l$. Note that a convention in literature is that $f > 0$ when mass flows inwards, whereas $g > 0$ when the viscous torque is outwardly positive, and we adopt that convention here. The binary torque is the sum of the advective and viscous torques at the inner boundary, $\dot J \equiv (f l - g)_{i=-1/2}$. In terms of the viscous torque $g_0$ and specific angular momentum $l_0$ at the center of the leftmost zone $i=0$, we arrive at an expression for the mass flow through the inner domain edge,
\begin{equation}
f_{-1/2} = \frac{g_0}{l_0 - \ell \Omega a^2} \, .
\label{eqn:torque_bc}
\end{equation}
Here we have approximated the derivatives in Eqn. \ref{eqn:rafikov_viscous_torque} to first order in the left half of the zone at $i=0$. The torque is reduced in magnitude over time by evolving $a$ according to Eqn. \ref{eqn:adot}. This fully describes the implementation of the inner boundary condition in the 1D code.

The surface density is initialized at $t=t_0$ to the approximate two-zone solution described in Sec. \ref{sec:model}; within $r < r_\nu(t_0)$ it is set to $\Sigma_{\rm in}(r, t_0)$ from Eqn. \ref{eqn:sigma_in}, and outside $r > r_\nu(t_0)$ it is set to $\Sigma_{\rm out}(r)$ in Eqn. \ref{eqn:sigma_out}. The outer boundary condition is enforced by assigning to the outermost zone interface the mass flux corresponding to the approximate two-zone solution. The calculations are evolved over some $10^5 t_{\rm dec}$ before and after the merger.

\subsection{The 2D Code}\label{subsec:2d_code}
The 2D code is \texttt{Sailfish}, a grid-based hydrodynamics code that solves the vertically-integrated Navier-Stokes equation in the time-dependent Newtonian potential of an orbiting binary \citep{zrake_sailfish_2024}. The binary components are on the grid, and accrete gas according to a ``sink'' prescription. We simulate the disk evolution some $10^4 t_{\rm dec}$ before and after the merger. During the inspiral phase $t<0$ the binary components move according to the Kepler two-body problem, modified to account for the GW driven orbital decay at the leading post-Newtonian order. The components merge at $t=0$. At $t>0$ the merger remnant is modeled by a single sink particle. This procedure is the same as in \cite{clyburn_dynamics_2024}. It is similar to what was done in \cite{krauth_disappearing_2023} except that here we have neglected a remnant recoil and the reduction of the central object's gravitating mass by GW radiation.

\begin{figure}
    \includegraphics[width=\columnwidth]{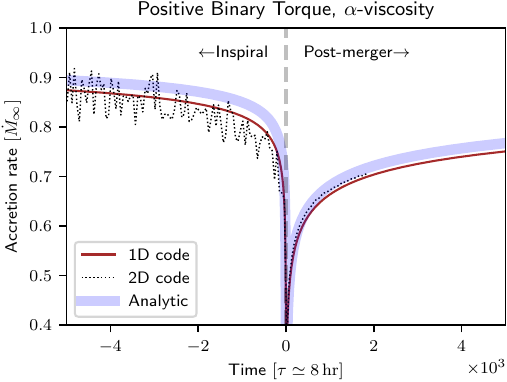}
    \caption{Predicted accretion rates for a positively torqued binary, over thousands of $\tau$ surrounding the merger, computed from the 1D code and 2D codes, and the analytic solutions in Eqns. \ref{eqn:mdot_pre} and \ref{eqn:mdot_post}. These calculations are for $n=1/2$. The 2D code uses a binary with $M_2/M_1=0.01$.}
    \label{fig:model_code_comparison_n5}
\end{figure}

For the new calculations presented here, the code uses the locally isothermal equation of state, where the gas temperature is an explicit function of space and time, and the sound speed is a small fraction $1/\mathcal{M}$ of the square root of the negative gravitational potential, $c_s = \sqrt{-\phi}/\mathcal{M}$. The constant $\mathcal{M}$ is called the orbital Mach number. Simulations are presented where either $\mathcal{M} = 10$ (positive torque runs) or $\mathcal{M} = 40$ (the negative torque run). For simulations done with the constant-$\nu$ ($n=0$) viscosity law, the initial conditions are the same as in \cite{tiede_gas-driven_2020}. For simulations done with the $\alpha$-viscosity ($n=1/2$) viscosity law, the initial conditions are the same as in \cite{clyburn_dynamics_2024}. More details about the 2D code can also be found in \cite{westernacher-schneider_multiband_2022}.

\subsection{Results}\label{subsec:results}
We present several examples of a three-way comparison between the accretion rates measured in the 1D and 2D codes, and the analytic $\dot M(t)$ in Eqns. \ref{eqn:mdot_pre} and \ref{eqn:mdot_post}. These comparisons enable us to empirically fix the value of $\kappa$, which was the only unknown parameter in the analytic model. The value $\kappa=5/3$ seems to give good overall agreement for both $n=0$ and $n=1/2$ viscosity models. The first two comparisons are for cases with positive binary torque ($\ell > 0$) and different viscosity models $n=0,1/2$. The third comparison is for a case with negative binary torque ($\ell < 0$). Note that in the 1D code $\ell$ is controlled directly via Eqn. \ref{eqn:torque_bc}, whereas in the 2D code $\ell$ is a consequence of the detailed disk-binary interaction and is sensitive to the binary mass ratio and the Mach number $\mathcal{M}$. To simulate a negatively torqued binary, the 2D the code was configured with an equal-mass binary and a relatively cold disk, $\mathcal{M}=40$.

\begin{figure}
    \includegraphics[width=\columnwidth]{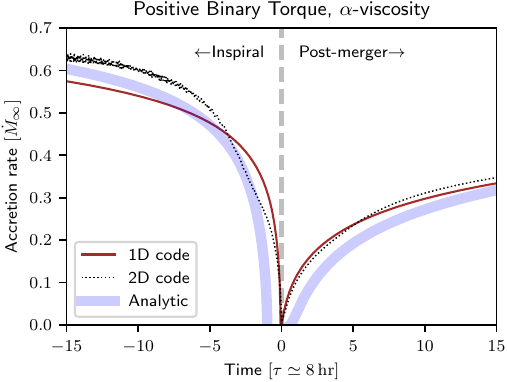}
    \caption{Same data as in Fig. \ref{fig:model_code_comparison_n5}, except zoomed in, showing $15 \tau$ before and after the binary merger.}
    \label{fig:model_code_comparison_n5_closeup}
\end{figure}

\subsubsection{Positive Binary Torque with $\alpha$-viscosity}\label{subsubsec:ell_pos_n5}
Figs. \ref{fig:model_code_comparison_n5} and \ref{fig:model_code_comparison_n5_closeup} show the three-way comparison for a case with $\alpha$-viscosity, $n=1/2$. For this comparison, the 2D code is run in a configuration with $M_2 / M_1 = 0.01$ and $\mathcal{M} = 10$, the same parameters as the fiducial model from \cite{clyburn_dynamics_2024}. When the mass ratio is small, the tidal torque can be neglected, and the angular momentum current is very nearly $\dot J = \dot M_2 \Omega a^2$, where $\dot M_2$ is the secondary's accretion rate. We find $\ell \simeq \dot M_2 / \dot M \simeq 0.74$. The time constant for this 2D calculation is then $\tau = \ell^{24/5} t_{\rm dec} \simeq 0.24 t_{\rm dec}$. For the fiducial disk viscosity (Sec. \ref{subsec:decoupling_timescale}) this corresponds to $\tau \simeq \unit[8]{hours}$; the horizontal span in Fig. \ref{fig:model_code_comparison_n5_closeup} thus covers about \unit[5]{days}.

For this setup, all three calculations lie within $10\%$ of one another throughout the inspiral and post-merger stages. Even within $15 \tau \simeq \unit[2.5]{days}$ of the merger, the agreement is still quite good. This might be surprising considering the model is not expected to be accurate where $|t| \lesssim t_{\rm dec}$.
The good agreement between these results in diverse numerical settings indicates that the time variation of the torque magnitude does indeed shape the disk structure and binary accretion rate around the time of a merger. It also suggests the analytic solution could be accurate enough for use in modeling the secular long-term AGN variability induced by binary mergers.

\subsubsection{Positive Binary Torque with Constant-$\nu$ Viscosity}
\begin{figure}
    \includegraphics[width=\columnwidth]{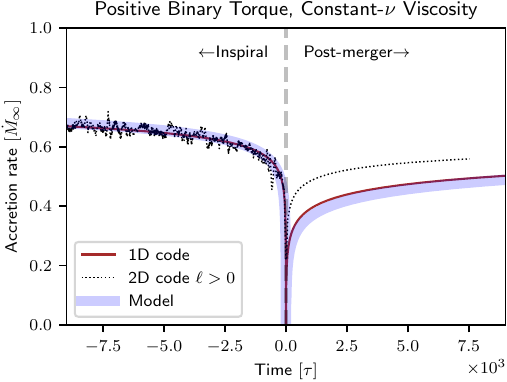}
    \caption{Accretion rates for a positively torqued binary over thousands of viscous decoupling times surrounding the merger, computed from the 1D and 2D codes, and Eqns. \ref{eqn:mdot_pre} and \ref{eqn:mdot_post}. These calculations are for $n=0$. In the 2D code the binary components are of equal mass and $\mathcal{M} = 10$.}
    \label{fig:model_code_comparison_n0}
\end{figure}

Fig. \ref{fig:model_code_comparison_n0} shows the three-way comparison for a constant-$\nu$ viscosity model, $n=0$. For this comparison, the 2D code is run in a configuration with an equal-mass binary, and $\mathcal{M}=10$, the same parameters as the ``standard'' binary accretion setup described in \cite{duffell_santa_2024}. It results in positive binary torque with $\ell \simeq 0.6 - 0.7$.

For this setup, all three methods to compute $\dot M$ are in close agreement throughout the inspiral stage. The analytic model and the 1D code are also in close agreement in the post-merger stage, but the result from the 2D code is systematically higher post-merger by roughly 20\% than the other calculations. This agreement indicates that the self-similar solution gives a good description of the pre- and post-merger evolutions of disks around binaries of different mass ratios and viscosity laws.

\subsubsection{Negative Binary Torque with Constant-$\nu$ Viscosity}
\begin{figure}
    \includegraphics[width=\columnwidth]{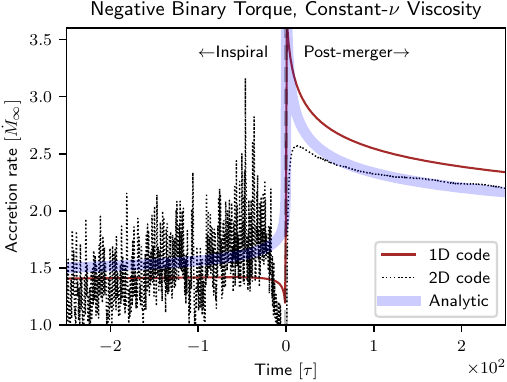}
    \caption{Accretion rates for a negatively torqued binary over hundreds of $\tau$ surrounding the merger, computed from the 1D and 2D codes, and Eqns. \ref{eqn:mdot_pre} and \ref{eqn:mdot_post}. These calculations are for $n=0$. In the 2D code the binary components are of equal mass and $\mathcal{M} = 40$.}
    \label{fig:model_code_comparison_negative_torque_mach40}
\end{figure}

Fig. \ref{fig:model_code_comparison_negative_torque_mach40} shows the three-way comparison for a negatively torqued binary, with constant-$\nu$ viscosity model, $n=0$. For this comparison, the 2D code is run in a configuration with an equal-mass binary, and $\mathcal{M}=40$, very similar to what was done in \cite{tiede_gas-driven_2020}. We measured the torque parameter in the 2D code to be $\ell \simeq -0.4$, consistent to within roughly 25\% with results from \citep{tiede_gas-driven_2020, penzlin_binary_2022}.

Aside from stochastic variation in the 2D code, all three calculations are in close agreement throughout the inspiral stage, showing an upward trending accretion rate. Note that Fig. \ref{fig:model_code_comparison_negative_torque_mach40} omits much of the pre-merger phase to focus on the shape of the accretion rates around the merger. However the analytic form of $\dot M_{\rm pre}(t)$ in Eqn. \ref{eqn:mdot_pre} blows up shortly before the merger at time $t = -\tau$. Around that time the 1D and 2D codes each show a reversal in the upward trending pre-merger accretion rate, and switch to the canonical downward-trend typically associated with viscous decoupling \cite[e.g.][]{dittmann_decoupling_2023}. The 1D and 2D accretion rates remain low until $t \lesssim 10 \tau$ after the merger, and then shoot back up again to meet or exceed the pre-merger maxima.

The true form of $\dot M$ around $t=0$ is surely sensitive to detailed physics, including radiation, magnetic fields, and relativistic gravity \cite[e.g.][]{williamson_binary_2022, most_magnetically_2024, avara_accretion_2024}, which are not accounted for here. We have checked whether the height of the accretion spike is sensitive to numerical parameters, and found that the height of the spike can be influenced by varying the size and mass subtraction rate parameters of the sink region. The data shown in Fig. \ref{fig:model_code_comparison_negative_torque_mach40} are a fair representation of several observed outcomes. A forthcoming study will give a detailed report on the the variety of possible outcomes in the $\ell < 0$ regime. In Sec. \ref{sec:observables} we report EM observables omitting the time window $|t| < \tau$. We believe the upward trending pre-merger accretion rate, and the gradual post-merger decline, are robust consequences of negatively torqued binary inspirals.

% However the 1D code and the analytic model both predict a large spike in the accretion rate at the time of the merger, which is not as pronounced in the 2D code. The 1D code and the analytic are predict systematically higher rates of accretion, by about 20\%, in the post-merger phase. We suggest this discrepancy might be the result of the 2D binary torque being only marginally negative. In a limit where $\ell \rightarrow 0$ the accretion curve, our physical picture implies the accretion rate should remain flat before and after the merger.

% Simulations with higher Mach number are numerically challenging, but are also very important as the disks around accreting massive black holes are believed to be extremely cold. The accretion spike at the time of the merger, and the subsequent long-lasting enhancement of the accretion rate, are both predicted by the model and the 1D code. Because these features could be telling electromagnetic signatures of a binary merger, it is important to numerically investigate in 2D or 3d simulations whether these enhancements become robust when the disk is colder, and the binary torque is more negative. It is worth noting that one simulation from \cite{farris_binary_2015} does show a significant accretion spike after the merger, however the binary torque was not reported, so we cannot say if that feature might have been a signature of the negative binary torque.

\section{EM Observables}\label{sec:observables}
Here we derive time-varying emission spectra from the analytic two-zone approximation of the disk evolution developed in Sec. \ref{sec:model}. The disk emission is approximated as a multi-temperature blackbody where the radial temperature profile $T(r)$ is determined by the local balance of viscous heating and thermal cooling. The procedure is the same as the one used in \cite{shapiro_filling_2010} to characterize the merger afterglow. In this section we only consider the case of an $\alpha$-viscosity law, however the light curves and emission spectra are not qualitatively changed by switching to a different viscosity law.

Our analytic model omits the complex gas flows in the near vicinity of the binary, within radius $r_\star(t)$, so it cannot directly constrain the form of the radiation emerging from that region. We show in Sec. \ref{subsec:missing_luminosity} that the fraction of the total accretion luminosity that is ``missing'' from our model can be reliably estimated, and we argue that it likely escapes as medium to hard X-ray photons. This would imply that the luminosity at optical through UV energies is dominated by the disk region outside $r_\star$, which is accurately described by the analytic model.

\subsection{The Decoupling Time}\label{subsec:decoupling_timescale}
The self-similar solutions developed in Sec. \ref{sec:model} vary in the parameter $t / \tau$, where $\tau \equiv |\ell|^{24/5} t_{\rm dec}$. The value of $t_{\rm dec}$ is obtained from  Eqn. \ref{eqn:adot} and the condition $r_\nu = r_{\rm dec}$,
\begin{eqnarray}
    t_{\rm dec} &=& \frac{5}{256} \frac{r_g}{\eta c} \left( \frac{128 \eta}{15 \bar \nu} \right)^{8/5} \\
    &\simeq& \unit[9.4]{day} \times \frac{M}{10^7 M_\odot} \left(\frac{\eta}{0.25}\right)^{3/5} \left(\frac{\bar \nu}{10^{-3}}\right)^{-8/5} \, ,
    \label{eqn:tdec}
\end{eqnarray}
where $\bar \nu \equiv \alpha / \mathcal{M}^2$. This corresponds to fiducial value of the decoupling radius of $r_{\rm dec} \equiv r_g (128 \eta / 15 \bar \nu)^{2/5} \simeq \unit[2.1]{AU}$. Note that this is much smaller than AGN disk sizes, inferred to be $\unit[10^{2-3}]{AU}$ \citep[e.g.][]{hawkins_timescale_2007, mudd_quasar_2018}. We have selected $\bar \nu = 10^{-3}$ to characterize the magnitude of the disk internal stresses. However this parameter introduces a considerable uncertainty; if $\bar \nu$ were even a factor of 10 smaller, then the decoupling time for an equal-mass binary would be lengthened to about a year. The sensitivity to $\eta$ is modest, for example the fiducial decoupling time is reduced to \unit[33]{hr} if the mass ratio is $M_2/M_1 = 0.01$. If the torque parameter were modest, say $\ell = 0.5$, then $\tau \equiv \ell^{24/5} t_{\rm dec}$ is about \unit[8]{hours}. If $\ell$ were large then $\tau$ could be an inaccessibly long timescale.

Expressions equivalent to Eqn. \ref{eqn:tdec} can be found in \cite{liu_doubledouble_2003, milosavljevic_afterglow_2005}. Those papers adopted smaller values for the effective disk viscosity $\bar \nu$, and arrived at significantly longer decoupling timescales of years to even millenia. Values of $t_{\rm dec}$ in the range of hours to weeks are implied by the larger values of $\bar \nu$ used in recent simulation studies including \cite{krauth_disappearing_2023} and \cite{clyburn_dynamics_2024}.
Given the uncertainties in $\bar \nu$ we keep in mind that AGN secular changes could happen over much longer or shorter timescales than our fiducial $t_{\rm dec} \sim \unit[]{days}$. If high-cadence multi-band light curves of an AGN around a binary merger are obtained, it might then be possible for $\tau$ to be obtained empirically. If the gas-induced torque magnitude can be extracted from GW measurements, that would correspond to a uniquely EM+GW probe of the disk internal stress.

\subsection{Disk Emission}\label{subsec:disk_emission}
The model surface density profile worked out in Sec. \ref{sec:model} is summarized here for the special case of an $\alpha$-viscosity prescription,
\begin{equation*}
	\Sigma(r, t) = \frac{1}{3 \pi \nu(r)}
    \begin{cases}
      \dot M_\infty       \left[ 1 - \ell \left( \frac{r}{r_{\rm dec}} \right)^{-5/16} \right]^{8/5} & r > r_\nu(t) \\
      \dot M_{\rm pre}(t) \left[ 1 - \ell \sqrt{\frac{a(t)}{r}} \right] & r < r_\nu(t) \, \, \& \, \, t < 0 \\
      \dot M_{\rm post}(t)  & r < r_\nu(t) \, \, \& \, \, t > 0 \, .
    \end{cases}
\end{equation*}
This function is piecewise continuous, and invokes Eqn. \ref{eqn:sigma_in} for $\Sigma_{\rm in}(r, t)$ and Eqn. \ref{eqn:sigma_out} for $\Sigma_{\rm out}(r)$. $\dot M_{\rm pre}(t)$ and $\dot M_{\rm post}(t)$ are given in Eqns. \ref{eqn:mdot_pre} and \ref{eqn:mdot_post} respectively. The radius of influence is
\begin{equation*}
 r_\nu(t) = 
    \begin{cases}
      r_{\rm dec} \left[ \frac{a(t)}{r_{\rm dec}} \right]^{8/3} & t < 0 \\
      \left[ \frac{6 \kappa \nu(r_{\rm dec}) t}{r_{\rm dec}^{1 / 2} } \right]^{2 / 3} & t > 0 \, .
    \end{cases}
\end{equation*}

The local rate of viscous energy dissipation, vertically integrated above the disk mid-plane, is
\begin{equation*}
    D(r, t) = \frac{9}{8} \nu \Sigma \Omega_{\rm kep}^2 \, .
\end{equation*}
The temperature of the disk photosphere is obtained from the balance of viscous heating and radiative cooling, $T(r, t) = [D(r, t) / \sigma]^{1/4}$. The power radiated from each face of the disk, per frequency interval, is denoted $L_\nu(t)$. It is obtained by numerically integrating the Planck function at frequency $\nu$ outwards from the disk inner edge,
\begin{equation*}
    L_\nu(t) = \int_{r_\star}^\infty \frac{2 h \nu^3}{c^2} \frac{1}{e^{h\nu / k T(r, t)} - 1} \, \pi dA \, .
\end{equation*}
The single-sided disk luminosity, $L_{\rm disk}(t)$, is obtained by integrating $L_\nu(t)$ over frequencies.

\subsection{Missing Accretion Luminosity}\label{subsec:missing_luminosity}
Our model gives a reliable description of the electromagnetic emission escaping from the disk before and after the merger, and likely captures the spectrum of the overall accretion system at optical through UV energies. However our model leaves out the complex gas flows in the near vicinity of the binary, and a majority of the overall accretion luminosity in the pre-merger phase is radiated from this region, especially from the circum-component disks. This missing accretion luminosity is equal to the difference between the specific orbital energy of gas at the inner edge $r_\star$ of the circumbinary disk, and that of gas at the innermost stable circular orbits $r_{\rm isco,1}$ or $r_{\rm isco,2}$ of the black holes, multiplied by the respective accretion rates,
\begin{eqnarray*}
	L_{\rm missing}
	&=& -\frac{G M \dot M}{2 r_\star} + \frac{G M_1 \dot M_1}{2 r_{\rm isco,1}} + \frac{G M_2 \dot M_2}{2 r_{\rm isco,2}} \\
	&\simeq& \frac{G M \dot M}{2 r_\star} \left(\frac{r_\star c^2}{6 G M} - 1\right) \, .
\end{eqnarray*}
The second approximation is exact for non-rotating holes, but otherwise introduces a small error, which depends on the preferential accretion parameter $\dot M_2 / \dot M_1$.

In what follows, we make the following additional approximations: (1) the flow is radiatively efficient down to the innermost stable circular orbits of the black holes, and (2) the missing accretion luminosity is radiated away at or above roughly $\unit[0.2]{keV}$. This assumption seems to be justified whether the binary mass ratio is small or large. When $M_2/M_1 \gtrsim 0.1$, the in-falling gas streams collide with the circum-component disks, and radiate their kinetic energy at or well above \unit[1]{keV} via shock heating \citep{farris_binary_2015, shi_how_2016, westernacher-schneider_multiband_2022, krauth_disappearing_2023, delaurentiis_relativistic_2024}. For small and intermediate mass ratios, most of the mass reaching the binary falls onto the smaller black hole \citep{lubow_disk_1999, munoz_circumbinary_2020, duffell_circumbinary_2020, siwek_preferential_2023}. For secondary black holes in the range of $10^{4-6} M_\odot$ the circum-secondary disk then radiates most of the missing power at around keV energies \citep{clyburn_dynamics_2024}.

\subsection{Spectral Energy Distribution}
\begin{figure}
    \includegraphics[width=\columnwidth]{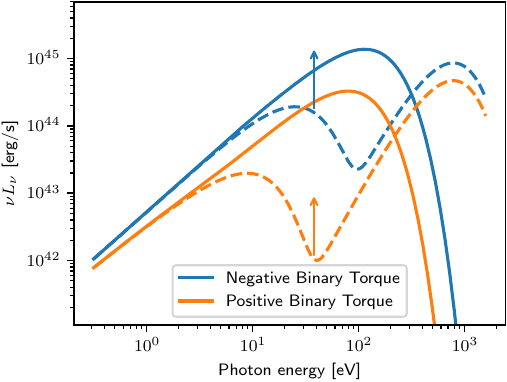}
    \caption{Disk emission spectra for positively vs. negatively torqued binaries. Dashed lines show the spectrum one year before the merger, and solid lines show the spectrum one year after the merger (arrows indicate the direction of the spectrum change).}
    \label{fig:disk_spectrum_different_ell}
\end{figure}

Fig. \ref{fig:disk_spectrum_different_ell} shows sample emission spectra at representative times, one year pre-merger and one year post-merger, for one case in which the binary is torqued positively by gas, and a second case in which it is torqued negatively. We assume an equal-mass binary with $M = 10^7 M_\odot$ and an $\alpha$-viscosity prescription, $\nu(r) = \bar \nu \sqrt{G M r}$ with $\bar \nu = 10^{-3}$. The inflow rate $\dot M_\infty$ is set to the Eddington rate, and the pre-merger missing accretion luminosity escapes as a single blackbody with temperature $\unit[0.2]{keV}$. Similar plots are easily made for binaries with different masses. As the black hole mass is increased, the spectrum is shifted horizontally to the left, as $M^{1/4}$, and vertically upwards in linear proportion to $M$.

The notch in the pre-merger spectral energy distributions are created by the distinct emission components: the circumbinary disk, whose emission peaks in the UV, and the missing accretion luminosity, characterized here as peaking in soft X-rays. Selecting a higher characteristic temperature for the gas in the near vicinity of the binary would shift the X-ray peak to the right, and would deepen the notch. The notch is reported by other authors as well: by \cite{kocsis_gas_2012-1} based on a 1D disk model, by \cite{shi_how_2016} based on 3D magnetohydrodynamics, by \cite{tang_late_2018} based on 2D hydrodynamics with a realistic equation of state, and by \cite{clyburn_dynamics_2024} based on 2D isothermal hydrodynamics with a toy emission model. Note that \cite{kocsis_gas_2012-1} solves an energy equation that includes a tidal heating term implied by negative binary torque, which we have not included in our estimated emission signatures. For $\ell \lesssim -1$ the tidal power $\sim \dot J \Omega$ the tidal heating is on the order of $r_g / a$  times smaller than $\dot M c^2$, so it could become noticeable late in the inspiral.

\begin{figure}
    \includegraphics[width=\columnwidth]{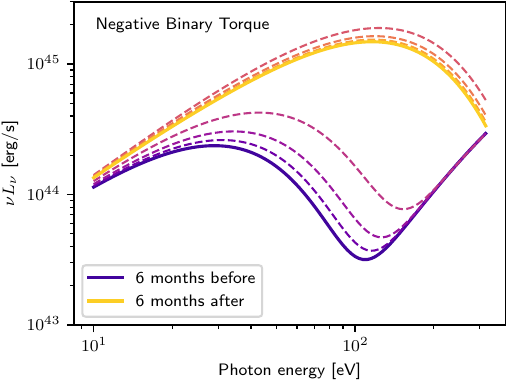}
    \caption{Time evolution of the disk emission spectrum for a negatively torqued binary. Solid lines show the spectrum 6 months before and 6 months after the merger, and the dashed lines show the spectrum in the intervening two-month intervals. None of the curves lie within the nominal decoupling time $t_{\rm dec} = \unit[9.4]{day}$ of the merger.}
    \label{fig:disk_spectral_evolution}
\end{figure}

The changes of the disk spectral characteristics in the years surrounding the merger are actually not very sensitive to the sign of the binary torque; either way the UV brightness gradually increases until the merger, and then the X-rays completely disappear. A more detailed view of the spectral evolution for the case of a negatively torqued binary is shown in Fig. \ref{fig:disk_spectral_evolution}. The notch grows shallower approaching the merger, and then disappears. The soft X-ray flare is evident as the spectrum in the sub-keV energy range is rapidly increased at the merger time, and then gradually reduced. Some of these features are easier to illustrate with multi-band light curves, which we look at next.

\subsection{Light Curves}
We show in Figs. \ref{fig:light_curve_positive_torque} and \ref{fig:light_curve_negative_torque} sample multi-band light curves in the decade surrounding the binary merger, for the cases of positive and negatively torqued binaries. The model parameters are the same as used in the previous sub-section. These light curves are computed assuming the missing accretion luminosity can be characterized as a single blackbody with temperature of $\unit[0.2]{keV}$. If the missing luminosity were to escape at higher energies, that would somewhat change the shape of the $\unit[1]{keV}$ light curves (green), but not the UV light curve (blue). The $\unit[0.2]{keV}$ light curve (orange) is somewhat affected in the case of positive torque, but in the case of the negative torque the $\unit[0.2]{keV}$ light curve is dominated by the inner portions of the disk, and not by the gas inside the low-density cavity.

A potentially exciting prediction of our model is the soft X-ray flare. It is simultaneous with the merger, occurring only when the torque is negative. It is shown as the orange curve in Fig. \ref{fig:light_curve_negative_torque}. The flare has a fast rise time on the order of a day, and declines much slower, over months to years. This prediction is not sensitive to the characterization of the missing accretion luminosity, provided the cavity temperature is higher than \unit[0.2]{keV}. The flare is powered by viscous dissipation and happens because of the cusp-like enhancement of the surface density at the inner edge of the circumbinary disk, as discussed in Sec. \ref{subsec:pre_merger_accretion_rate}. The cusp follows the binary inwards up until $t = -t_{\rm dec}$, and then promptly floods the merger remnant with gas. Note that the flood of gas begins at $t=0$ and peaks $\sim t_{\rm dec}$ as envisioned previously in \cite{liu_doubledouble_2003, milosavljevic_afterglow_2005, shapiro_filling_2010}. This is kind of obvious in hindsight. The cusp lies at radius $r_{\rm dec}$ at time $t_{\rm dec}$, and by definition moves inwards 

Regardless of the sign of the binary torque, our model predicts a gradual increase of the UV brightness throughout the inspiral. This effect comes from the increasing peak temperature of the circumbinary disk as it contracts in step with the binary orbital decay \citep{clyburn_dynamics_2024}. However in the pre-merger phase the positively torqued binary is about a factor of 10 dimmer in the UV than the negatively torqued binary. This is the signature of the gradual increase of the overall accretion luminosity predicted only when the torque is negative. That effect also implies that a positively torqued binary grows brighter in the UV by a factor of about five in the several years following the merger, whereas the negatively torqued binary becomes modestly dimmer over that same time frame. Note that in both scenarios the UV brightness approaches the same steady-state value $\sim \unit[10^{44}]{erg/s}$, but from below and above respectively.

The light curves at $\unit[1]{keV}$ confirm the disappearance of X-ray emission at the time of the merger first predicted in \cite{krauth_disappearing_2023}, and also confirmed in \cite{franchini_emission_2024, clyburn_dynamics_2024}. We find that when the torque is positive, the X-ray dimming is relatively smooth, and more gradual than when the torque is negative. In the case of negative torque (green curve in Fig. \ref{fig:light_curve_negative_torque}) the X-rays remain bright right up until the time of the merger, and then shut off practically as a step function. This finding needs to be directly confirmed in simulations with self-consistent thermodynamics and a negatively torqued binary. The disk conditions in \cite{krauth_self-lensing_2024} were rather warm, consistent with a positive binary torque and the smooth decline of the X-ray emission shown in their Fig. 4. Realistic disks are likely much colder and produce a negative binary torque.

\section{Discussion}\label{sec:discussion}
\subsection{Changing-Look Inspirals}
AGN which show significant lasting changes in their blue continuum and broad-line emissions are referred to as changing-look AGN \citep[see e.g.][for a recent review]{ricci_changing-look_2023}. A subset of changing-look events are caused by transient obscurations of the disk, corona, or broad-line emitting regions, however others are likely associated with abrupt changes in the luminosity of the accretion flow \citep[e.g.][]{yang_probing_2023}. These events are called changing-state (CS) AGN events. Our results indicate that abrupt and chromatic changes of AGN appearances are expected to accompany massive black hole binary inspirals. We call such events changing-look inspirals (CLI's), and we propose that some documented CS AGN events could be CLI's.

\begin{figure}
    \includegraphics[width=\columnwidth]{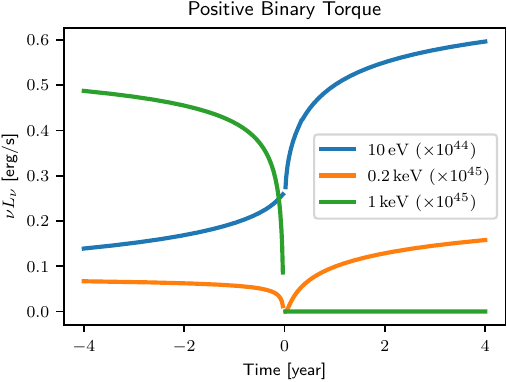}
    \caption{Light curves for a positively torqued binary, $\ell = 1$. The binary mass is $10^7 M_\odot$ and $t_{\rm dec} = \unit[9.4]{days}$.
    The pre-merger missing accretion luminosity escapes as a single blackbody with temperature $\unit[0.2]{keV}$. The time window within $t_{\rm dec} = \unit[9.4]{day}$ of the merger is omitted.
    }
    \label{fig:light_curve_positive_torque}
\end{figure}

\begin{figure}
    \includegraphics[width=\columnwidth]{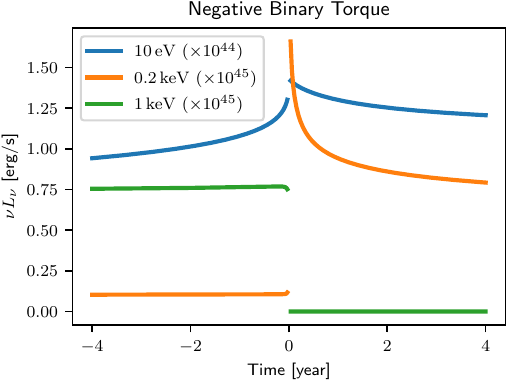}
    \caption{Lights curve for a negatively torqued binary, $\ell = -1$. The binary mass is $10^7 M_\odot$ and $t_{\rm dec} = \unit[9.4]{days}$.
    The pre-merger missing accretion luminosity escapes as a single blackbody with temperature $\unit[0.2]{keV}$. The time window within $t_{\rm dec} = \unit[9.4]{day}$ of the merger is omitted.
    }
    \label{fig:light_curve_negative_torque}
\end{figure}

Figs. \ref{fig:light_curve_positive_torque} and \ref{fig:light_curve_negative_torque} show examples of where the disk emission changes dramatically at the merger time. For a Type-A event (Fig. \ref{fig:light_curve_positive_torque}) the EM change includes a five-fold year-like increase in the disk brightness at \unit[10]{eV}, and a shut-off of the X-ray emission above \unit[1]{keV}. For a Type-B event (Fig. \ref{fig:light_curve_negative_torque}) the EM change may include a large-amplitude asymmetric sub-keV X-ray flare with very fast rise and months-like decay, and a long-lived 50\% increase in the disk UV brightness. Because the broad-line region is illuminated by the disk, such changes could correspond to significant alterations, including the simultaneous appearance or disappearance of the blue continuum and broad-line emissions. A rigorous test of a CLI hypothesis for documented CS AGN events will require fitting multi-wavelength data to detailed radiative transfer calculations. Such calculations could be carried out, using our model in Sec. \ref{subsec:disk_emission} for the variable disk emission as the light source.

Nevertheless there is a very well studied changing-look AGN, Mrk 1018, whose behavior seems broadly consistent with the signature of a Type-B CLI which might have taken place in the early 1980's. This system has undergone a full cycle of spectral changes, with broad-line emission features appearing in the mid to late 1980's \citep{cohen_variability_1986, goodrich_spectropolarimetry_1989} and then disappearing again sometime before 2016 \citep{mcelroy_close_2016, husemann_close_2016}. The emergence of broad-line emissions is thought to arise from an enhancement of the accretion disk luminosity \citep{veronese_interpreting_2024}, as we predict must accompany the post-merger gas flow in a Type-B event (Fig. \ref{fig:light_curve_negative_torque}). Mrk 1018 is hosted in a late-stage merger galaxy \cite{mcelroy_close_2016}, so a massive black hole binary has likely formed sometime in the last $\sim$Gyr. Our model for the gradual restructuring of an accretion disk around a merging binary black hole provides a rigorous basis for a detailed physical model to test a merger hypothesis for Mrk 1018.

The possibility of an orbiting secondary black hole in Mrk 1018 was proposed in \cite{mcelroy_close_2016, husemann_close_2016}, but a detailed model of the emission signature based on orbital motion was not put forward, and there is no conclusive evidence of more than one cycle of spectral variations this source. \cite{kim_recoiling_2018} has suggested the variations observed so far might correspond to an epicyclic oscillation of a recoiling black hole merger remnant, however VLBI measurements have not confirmed the implied proper motion \citep{walsh_vlbi_2023}. Future monitoring of this source should constrain the orbiting binary and recoiling / oscillating remnant scenarios.

Future LISA detections will trigger multi-wavelength followup monitoring campaigns to identify host galaxy candidates \cite[e.g.]{kocsis_premerger_2008}. We recommend to prioritize any AGN's in the LISA error volume which have shown a changing-state event, or which have exhibited secular variability in the preceding years. Even if there is no target of opportunity, AGN monitoring by optical and UV surveys might be sufficient to reveal temporally coincident CS-GW events. Given the relatively low rate of both CS events and LISA inspirals, the rate of chance coincidences might be vanishingly low.

\subsection{Merger Afterglows}
Merger afterglows on time scales of years to decades were discussed in \cite{milosavljevic_afterglow_2005} and in \cite{shapiro_filling_2010}, based on the viscous refilling of a depleted central cavity. Our prediction is qualitatively similar to theirs, but only for the case when the gas produces net-positive binary torque during the inspiral; compare for example our Fig. \ref{fig:light_curve_positive_torque} to Shapiro's Fig. 5. There is an apparent discrepancy between the afterglow temperature we predict, and that predicted in \cite{milosavljevic_afterglow_2005}, their prediction has the re-brightening phase being mainly in harder X-rays. The discrepancy seems to originate from their parameterization of disk surface density (their Eqn. 3), it yields a significantly higher temperature at the inner edge of the remnant disk than we find in our dynamical model. The afterglow temperature predicted in \cite{shapiro_filling_2010} is more in line with what we have predicted here, see his Fig. 4 and Eqns. 19 -- 21.

Viscous refilling of a depleted central cavity is also surmised to lead to a delayed jet revival, as gas from a hollowed-out disk eventually reaches the remnant \citep{liu_doubledouble_2003}. The delay time they derived is on the order of of $10^5$ years, much longer than the years-like re-brightening phase we have estimated here. The difference seems to originate in their adoption of a very low disk viscosity, and of an orbital hardening law (their Eqn. 3) that was based on the second type viscous migration of a low-mass secondary.

\subsection{Merger Flares}
The possibility of a bright flare roughly simultaneous with a massive black hole binary merger was proposed in \cite{armitage_accretion_2002}. They presented time-dependent 1D calculations of a disk surrounding a massive primary black hole, and embedding a lower mass secondary black hole. The secondary was characterized as exerting a retrograde tidal torque to the inner disk, strong enough to force the inner disk gas to accrete to the primary leading up to the merger. A similar mechanism, now sometimes called a snowplow, is also discussed in \cite{lodato_black_2009} and in \cite{chang_fossil_2010}. However \cite{baruteau_no_2012} demonstrated using a 2D hydrodynamics calculation that the snowplow is likely an artifact of 1D disk prescriptions that forbid gas from crossing the low-density gap opened by the secondary. \cite{clyburn_dynamics_2024} showed that to squeeze the inner disk at the rate of the GW inspiral would require a large (and \emph{positive}) binary torque. Thus if an X-ray flare is observed around the time of a massive black hole binary merger, a GW measurement revealing a negative gas-induced torque could rule out a snowplow. Similarly a positive gas-induced torque could rule out the weakening-torque effect. Also note that a snowplow effect is expected to produce a flare shortly before the merger, whereas the weakening (negative) torque effect produces a flare after the merger.

Post-merger accretion spikes seem not to have been seen so far in multi-dimensional hydrodynamics simulations. We think this is because published simulations are yet of rather warm disks, with orbital Mach numbers not much greater than $\mathcal{M} = 10$, and such disks generally produce a net-positive binary torque, and the associated downward-trending pre-merger accretion rate. Colder disks on the other hand generally produce more negative torques \citep{tiede_gas-driven_2020, penzlin_binary_2022}, because streamline intersections needed to capture gas parcels onto single black hole orbits can be suppressed when the in-falling gas streams are very cold \citep{tiede_suppressed_2024}. Even though colder disks are more challenging to simulate, they are probably more representative of black hole accretion disks in nature \citep[e.g.][]{shakura_black_1973}.

\subsection{Suppressed Accretion in Binary Black Holes}
Some recent studies \citep{ragusa_suppression_2016, dittmann_survey_2022, tiede_suppressed_2024} find that when circumbinary disks are very cold, there can be a suppressed efficiency of gas capture onto the black holes. If massive black hole binaries generically interact with gas from circumbinary disks, but do not form minidisks, there are many interesting implications, such as the possibility mentioned in \cite{tiede_suppressed_2024} that photometrically peculiar high-redshift galaxies seen in JWST deep fields, so-called little red dots \citep{matthee_little_2024}, could be non-accreting binary AGN's. It would also imply that AGN's do in fact ``turn on'' (at least in optical through X-ray light) following the merger, as envisioned in e.g. \cite{shapiro_filling_2010} and \cite{tanaka_witnessing_2010}. The time-dependent disk solutions we developed in this paper would need to be adapted to accommodate regimes of suppressed binary accretion, and might reveal distinct shapes of the post-merger light curves, relative to what was found in the aforementioned 2010 studies, which did not account for the history of the binary contraction in choosing an initial condition for the circum-remnant disk.

\subsection{Measuring the Gas Torque from GW's}
Can the gas-induced torque be measured? There is a lot of work done to understand how environmental factors perturb gravitational wave forms \citep[e.g.][]{chakrabarti_gravitational_1996, levin_starbursts_2007, kocsis_observable_2011, zwick_priorities_2023}. However the inverse problem might be much harder because of parameter degeneracies. An example, is that an accelerated inspiral does not require a negative torque, because of the mass change effect. The mass change effect is seen in this equation,
\begin{equation}
\dot \Omega_{\rm gas} = \Omega \left(5 \frac{\dot M}{M} + 3 \frac{\dot \eta}{\eta} - 3 \frac{\dot J}{J}\right) \, ,
\label{eqn:omega_dot_gas}
\end{equation}
for the gas (non-GR) contribution to the time derivative of the binary orbital phase. This equation comes from writing the orbital phase as $\Omega = G^2 M^5 \eta^3 / J^3$. Integrating it twice in time gives half the accumulated GW phase drift. There are studies that set the phase drift in direct proportion to the gas torque \citep[e.g.][]{derdzinski_probing_2019, derdzinski_evolution_2021, garg_imprint_2022, garg_measuring_2024, dittmann_decoupling_2023}, so do not include the mass change effect.

The ansatz $\dot J = \ell \dot M \Omega a^2$ has been used throughout the paper. When the mass ratio is large, $\eta \simeq 1/4$ and the $\dot \eta$ term can be neglected. Eqn. \ref{eqn:omega_dot_gas} reduces to
\begin{equation*}
\dot \Omega_{\rm gas} \simeq \Omega \left(5 - 12 \ell\right) \frac{\dot M}{M} \, .
\end{equation*}
Thus for a large mass ratio inspiral, an accelerated inspiral means that $\ell$ was less than $5/12$. Whereas, for a low mass ratio inspiral the accretion favors the secondary, $\dot M_2 / M_2 \gg \dot M_1 / M_1$, and $\dot \eta \rightarrow \dot M / M$. Eqn. \ref{eqn:omega_dot_gas} becomes
\begin{equation*}
\dot \Omega_{\rm gas} \simeq \Omega \left(1 - 3 \ell \right) \frac{\dot M}{M_2} \, .
\end{equation*}
Thus for an intermediate or extreme mass ratio inspiral, an accelerated inspiral means that $\ell$ was less than $1/3$. We can still predict that Type-B events will correlate with accelerated inspirals.

More generally, $\ell$ might be ``parameter-estimated'' by matching observations against waveform templates, derived from the scenario discussed in this paper. A full GW template would be implied at leading post-Newtonian order by Eqns. \ref{eqn:mdot_pre} and \ref{eqn:omega_dot_gas}, given prior predictions of $\dot \eta$. These could be obtained from simulations.

Eqn. \ref{eqn:omega_dot_gas} illustrates that to measure $\ell$ based on a GW phase drift and our model for $\dot M_{\rm pre}$ requires a prior on the large-scale inflow rate $\dot M_\infty / M$. Multi-wavelength EM monitoring can constrain the quiescent accretion rate $\dot M_\infty$. The remnant mass can be constrained from the GW chirp mass, or from late-time iron-line emission from the post-merger disk. The instantaneous accretion rate $\dot M(t)$ is also the rate of binary mass increase, and it should cause a gas-induced ``amplitude drift'' analogous to the phase drift. Gas effects such as winds \citep{kocsis_observable_2011}, stochastic forcing \citep{zwick_dirty_2022}, and higher-order relativistic effects including apsidal precession might be hard to disentangle \citep{tiede_disk-induced_2024, delaurentiis_relativistic_2024}.

\section{Summary}\label{sec:summary}
We have described a robust physical basis for a ``changing-look inspiral'' (CLI), in which an AGN undergoes a lasting change of appearance in connection with the GW driven inspiral of a central massive black hole binary. CLI's might account for a subset of documented changing-state AGN events, in which an abrupt appearance or disappearance of broad-line spectral features is attributed to a restructing of the inner portions of the accretion disk.
The main take-aways of our paper are summarized here:
\begin{enumerate}
    \item CLI's have two sub-types: In Type-A (attenuated) events the gas torques on the binary are positive, and the AGN is dimmest at the merger time. In Type-B (brightening) events the gas torques are negative and the AGN is brightest and may produce a significant soft X-ray flare shortly following the merger. In both Type-A and Type-B events there is a long-lasting chromatic change in the AGN appearance around the time of the merger. We derived a model for the AGN disk (blue continuum) emission, which could be readily incorporated into a full radiative transfer calculation of the variable AGN emission produced in the time period including a massive black hole binary merger.
    \item The secular variability of the AGN disk emission stems from a generic behavior of disks around contracting binaries, which we called a ``weakening-torque effect''. It shapes the radial mass distribution outside the binary orbit, and in turn modulates the flow of mass to the binary during the inspiral, and to the black hole remnant following the merger. The effect can be seen in a model problem wherein an axisymmetric thin disk is subjected at its inner edge to a time-varying torque. We showed three-way quantitative agreements in the evolution of the disk structure computed from 2D grid-based hydrodynamics simulations, 1D time integrations of the standard disk equation, and an approximate two-zone model which yields self-similar time-dependent solutions. These agreements were seen in three representative disk-binary configurations with various gas temperatures, viscosity laws, and mass ratios.
    \item We discussed the potential for joint EM+GW observations of massive black hole binary inspirals to corroborate our prediction of a dichotomy of AGN variability patterns associated with binary mergers. We pointed out that the sign of the gas-induced binary torque need not correspond to the direction of the phase drifting of the GW signal. We predict that if a merger X-ray flare is observed, then the phase drift should be prograde, whereas for dimming events the phase drift could be prograde or retrograde. Parameter estimation applied to the sign of the gas-induced binary torque may require the application of physically motivated GW templates, and such templates can be constructed from Eqn. \ref{eqn:mdot_pre}, together with priors on the preferential accretion parameter derived from simulations.
    \item Future LISA detections will trigger multi-wavelength followup monitoring campaigns to identify to host galaxy. We recommend to prioritize any AGN's in the LISA error volume which have shown a changing-state event, or which have exhibited chromatic brightness trends in the preceding years, especially a gradual UV brightening, or a trend in either direction of the X-ray brightness.
\end{enumerate}

The biggest uncertainties in our predictions are the magnitudes of the disk internal stress $\bar \nu$ and the torque parameter $\ell$. If disks around black hole binaries have much longer viscous timescales than those we adopted in this paper, or if they operate in a regime of low accretion efficiency (very negative $\ell$), then the secular AGN variability from a binary merger could operate over inaccessibly long timescales. Well-sampled light curves of a CLI (Type-A or Type-B) could be fit to obtain the characteristic timescale $\tau$, which depends only on $\ell$ and $\bar \nu$. LISA measurements that constrain the gas-induced torque would then simultaneously constrain the magnitude of the disk internal stress.

\section*{Acknowledgements}
J. Zrake acknowledges support from the LISA Preparatory Science Program (LPS) through NASA Award No. 80-NSSC-24K0440.
M. Clyburn acknowledges support from the NASA Future Investigators Program (FINESST) through Award No. 80-NSSC-23K1443.
The authors are grateful to Christopher Tiede and Zoltan Haiman for insightful comments.
All simulations were performed on Clemson University's Palmetto cluster.

%%%%%%%%%%%%%%%%%%%%%%%%%%%%%%%%%%%%%%%%%%%%%%%%%%
\section*{Data Availability}

The data underlying this article will be shared on reasonable
request to the corresponding author.



\bibliographystyle{mnras}
% \bibliography{references}

%\begin{thebibliography}{99}
%\end{thebibliography}

%%%%%%%%%%%%%%%%%%%%%%%%%%%%%%%%%%%%%%%%%%%%%%%%%%

%%%%%%%%%%%%%%%%% APPENDICES %%%%%%%%%%%%%%%%%%%%%

%\appendix


%%%%%%%%%%%%%%%%%%%%%%%%%%%%%%%%%%%%%%%%%%%%%%%%%%

% Don't change these lines
\bsp	% typesetting comment
\label{lastpage}
\end{document}